\documentstyle[aps,eqsecnum]{revtex}
\begin{document}
\draft
\preprint{\vbox{\hbox{UT-682} \hbox{SNUTP 94-55}
\hbox{hep-th/9406192} \hbox{June 1994} }}
\title{Collective Field Theory of the Fractional Quantum Hall Edge State
        \\ and the Calogero-Sutherland Model}
\author{Satoshi Iso${}^1$ and Soo-Jong Rey${}^2$}
\address{Department of Physics, University of Tokyo\\
         Bunkyo-ku, Tokyo 113 Japan${}^1$ \\
    Physics Department and Center for Theoretical Physics \\
    Seoul National University, Seoul 151-752 Korea${}^2$ \\
}
\date{\today}
\maketitle

\begin{abstract}
\noindent  Using hydrodynamic collective field theory approach we show that
one-particle density matrix of the $\nu=1/m$ fractional quantum Hall edge
state interpolates between chiral Luttinger liquid behavior $\langle
\psi^{\dagger}(r) \psi(0) \rangle \sim r^{-m} $ and  Calogero-Sutherland model
behavior $\langle \psi^{\dagger}(r) \psi(0) \rangle  \sim r^{-(m+1/m)/2} $
as the droplet width is varied continuously. Low-energy excitations are
described by $c=1$ conformal field theory of a compact boson of radius
$\sqrt m$.
The result suggests complementary relation between the two-dimensional
quantum Hall droplet and the one-dimensional Calogero-Sutherland model.
\end{abstract}
\pacs{73.40.H, 05.30.-d}
Recently there has been renewed interest to the Calogero-Sutherland (CS)
 model of one-dimensional (1-d) fermions with long-range
interactions~\cite{calogero,sutherland} as an exactly soluble Luttinger
liquid. The model may be viewed as an ideal gas of anyons as the interaction
is purely statistical~\cite{anyon,haldane}. In addition dynamical correlation
functions of the model have been calculated~\cite{haldane,simon,poly}.
The CS model also exhibits many similarities with the fractional quantum
Hall effect (FQHE) such as the fractional statistics properties of the
quasi-holes~\cite{haldane}, hierarchical extensions~\cite{kawakami} and
$W_{\infty}$ symmetry~\cite{W,wadati}. As such one naturally inquire whether
there exists any deep relation between the FQHE and the CS model.
\par
The FQHE is two-dimensional while the CS model is a 1-d system. Nevertheless
it is possible to show that the FQHE restricted to the lowest Landau level
(LLL) is directly related to the CS model. For a high external magnetic field
$B$ the FQHE system undergoes a `dimensional reduction' so that the LLL wave
function may be interpreted as wave function of a 1-d system~\cite{iso}.
Noting that the commutation relations of the guiding center coordinates
$[X,Y]= i /B $ 1-d interpretation of the LLL wave functions is obtained in
the basis $|x \rangle$ of the LLL:
$ X|x \rangle =x|x \rangle$, $ Y|x \rangle = p_x /B|x \rangle$.
In this representation the $X$ coordinate is diagonalized while the $Y$
coordinate is interpreted as the conjugate momentum to $X$. In addition
operators acting on the LLL can be interpreted as the operators of the
1-d fermion system. Therefore the FQHE droplet may be viewed as occupied
phase space of the 1-d fermions.
\par
Using the 1-d representation of the LLL wave functions, the Laughlin
state on a cylinder of circumference $L$ is expressed as~\cite{iso}
\begin{eqnarray}
&& \langle s_1 \cdot \cdot \cdot s_N|\Psi \rangle \nonumber \\
&& =
e^{-{\hbar \over 2B} \sum_i \left( {\partial \over \partial s_i} \right)^2}
\prod_{i<j} (e^{i2\pi s_i/L} -e^{i2\pi s_j/L} )^m.
\label{Laughlin}
\end{eqnarray}
Width of the  FQH droplet is $\delta y =2 \pi\hbar N m /B L = 2m \hbar k_F
/B $ where $k_F=\pi N/L$.
In the large $B$ limit, the droplet becomes narrow ($\delta y
\rightarrow 0$) and the Laughlin state coincides exactly with the
ground state of the Sutherlad model. (The total momentum is shifted to
zero in the definition of the above Laughlin state.)
\par
Long distance behavior of the fermion one-particle density matrix for
the Laughlin state Eq.~(\ref{Laughlin}) is controlled by the boundary
excitations of the FQHE since there are no other gapless excitations in
the bulk. On general ground Wen \cite{wen} obtained the edge state
correlation function: $\langle \psi^{\dagger}(r) \psi(0) \rangle \sim r^{-m}$
where the filling factor $\nu$ is $\nu=1/m$.
\par
On the other hand the fermion correlation for the CS model is given
by~\cite{kawakamiyang}
$\langle \psi^{\dagger}(r) \psi(0) \rangle \sim r^{-(m+1/m)/2} $.
Since the Laughlin state coincides with the ground state of the CS
model in the narrow channel limit the fermion correlation for edge state
of the FQHE system must interpolate its behavior from $r^{-m}$ to
$r^{-(m+1/m)/2}$ when the width of the FQHE droplet is adiabatically
narrowed by changing either the strength of the mangetic field or the electron
number density. Consistency of the existence of two different exponents
$m$ and $(m+1/m)/2$ have been discussed in~\cite{kawakami} based on chiral
constraints imposed on the CS model and in~\cite{stonefisher} by identifying
`dressed' fermion operator in the effective thoery of the FQHE.
\par
In this letter we show explicitly that the correlation exponent interpolates as
anticipated above. Our starting point is the hydrodynamic description of
the FQH droplet dynamics using collective field theory. It is a straightforward
generalization of the integer QHE case~\cite{iks}.
\par
It has been known that the $\nu = 1/ m$ FQHE may be interpreted as arising
due to condensation of bosonized fermion $\psi$ once appropriate
flux of Chern-Simons gauge field is attached to it~\cite{girvin}.
The Lagrangian reads
\begin{equation}
 {\cal L} = \bar\psi \Pi_0\psi -{1\over{2m_0}}\bar\psi (\Pi^x +i\Pi^y )
(\Pi^x -i\Pi^y )\psi +{\cal L}_{CS}
\end{equation}
where $m_0$ is the fermion mass, $ \Pi_{0} = i\partial _0 - A_{0} - a_{0},
\Pi_{i} = -i\partial _i - A_{i} - a_{i}, i=1,2$ and $ {\cal L}_{CS} =
-(1 / 4\pi m) \epsilon^{\mu\nu\rho}a_{\mu}\partial_{\nu}a_{\rho}$.
External magnetic field and mechanical potential are described by
${\bf A}$ ($ - B = \nabla \times {\bf A} < 0$) and $A_0(x,y)$ respectively.
In a strong magnetic field the low-energy excitations are restricted to
the LLL. Projection to the LLL is made by taking $m_0 \rightarrow 0$ and
yields~\cite{gjgmp}
\begin{equation}
        (\Pi_x -i\Pi_y )\psi =0.
        \label{LLL}
\end{equation}
Next we change the bosonized fermion fields $\psi$, $\bar\psi$ to hydrodynamic
fields $\rho$, $\theta$ through ${\psi =\sqrt{\rho} e ^{i\theta}}.$
The phase field $\theta$ consists of a regular part and a singular part:
$\theta = \theta_{\rm reg} +\theta_{\rm sing}$.
The singular part originates from vortex configurations:
\begin{equation}
\partial^{\mu}\theta_{\rm sing} = v^{\mu},\ \ \  \nabla\times{\bf v} =
2\pi\rho_V, \ \ \ \nabla\cdot{\bf v}=0
\label{singular}
\end{equation}
where $\rho_V$ denotes density of the vortices.
Acting $(\partial_x + i \partial_y)$ on the LLL condition Eq.~(\ref{LLL})
yields two constraint equations in the gauge $\nabla \cdot
{\bf A} = 0$
\begin{eqnarray}
 && \nabla\times{\bf v}  +\nabla\times{\bf A}
+\nabla\times{\bf a}- {1\over 2}\nabla^2\ln\rho =0, \label{LLL1} \\
&& \nabla ^{2} \theta _{reg} + \nabla\cdot {\bf a}=0. \label{LLL2}
\end{eqnarray}
\par
Imposing Eq.~(\ref{LLL}) the Lagrangian is written as
\begin{equation}
L = - \int d^2 x \ [ \rho (\dot{\theta} + A_0)
+{1\over{4\pi m}}\epsilon^{ij}a_{i} \dot{a_{j}}]
 \label{Lag}
\end{equation}
where we have integrated over $a_0$ and obtained the constaint
$\nabla\times{\bf a}=2 \pi \rho m$.  Eq.~(\ref{LLL2})
implies that the regular part of the phase in Eq.~(\ref{Lag}) cancels
the last term. Therefore the reduced Lagrangian of the FQH droplet is
\begin{equation}
{\cal L} = -(v_0 + A_0 )\rho
\label{lagran}
\end{equation}
supplemented by the subsidiary condition
\begin{equation}
B + {1\over 2}\nabla^2\ln\rho -2\pi\rho_V
-2\pi\rho m =0.
\label{subsidiary}
\end{equation}

To investigate low-energy, gapless excitations of the FQH droplet
we now quantize the hydrodynamic collective field theory derived above.
Ground state of the droplet is determined by the external
potential $A_0$. Following~\cite{iks} we choose the potential as
$A_0(x, y) = B^2 y^2 / 2 + v(x) - \mu$.
In the large magnetic field limit Eq.~(\ref{subsidiary}) is solved
approximately by
\begin{equation}
\rho(x, y, t) = {B \over 2 \pi m}\ \theta (y_+ (x, t) - y) \ \theta
(y - y_-(x, t)).
\label{density}
\end{equation}
Boundary shape of the static ground state droplet depends on the 1-d
potential $v(x)$.
The dynamical degrees of freedom are the droplet boundaries
$y_\pm(x,t)$ fluctuating near the ground state configuration.
\par
Solving Eq.~(\ref{singular}) for $\theta_{\rm sing}$ the first term
of Eq.~(\ref{lagran}) reads
\begin{equation}
L_0 = - \! \! \int \! d^2 x \ v_0\rho =
2 \pi \! \int\!\!\!\int\! \rho (x) G(x-x^{\prime}) \dot{\rho}_V (x^{\prime})
d^2 x d^2 x'
\end{equation}
where $G$ is the Green's function satisfying  $\nabla^2 G = 0$,
$\nabla\!  \times \! \nabla G(x-x') = \delta^2 (x-x').$
Using Eq.~(\ref{density}) $L_0$ becomes
\begin{eqnarray}
L_0 &=& {B^2 \over 4 \pi m} \int \! d^2 x \!\! \int \! d^2 x' \
\theta (x - x')
\nonumber \\
 &\times& [y_+(x) \dot{y_+}(x') - y_-(x) \dot{y_-}(x')] .
\label{quadL}
\end{eqnarray}
 From $L_0$ we obtain commutation relations for $\rho(x)$, $\rho_V(x)$
and, in turn, $y_\pm (x,t)$
\begin{equation}
[ y_\pm(x), y_\pm(x')]_{\rm ET} = \mp i {2 \pi m \over B^2}
\delta ' (x - x').
\label{commrel}
\end{equation}
Physically $y_\pm (x, t)$ generate the left and the right moving
chiral edge state excitations respectively.
\par
The second term of Eq.~(\ref{lagran}) is the Hamiltonian
\begin{eqnarray}
H &=& \int \rho A_0 \nonumber \\
&=& {B \over 2\pi m} \int dx
[ {B^2 \over 6}\left( y_{+} ^3 (x) - y_{-}^3 (x)\right)
\nonumber \\
    &+&(v(x)-\mu) \left( y_{+} (x)- y_{-}(x)\right) ].
\label{hamiltonian}
\end{eqnarray}
For small amplitude the Hamiltonian measures elastic curvature energy of
$y_\pm(x,t)$.
\par
 From the above collective field theory we now construct operators
for low-energy quasi-particle excitations along the boundaries of
the FHQ droplet. This is most conveniently described in terms of the left-
and right-moving chiral boson fields $\varphi_{\pm}$
 \begin{equation}
y_{\pm}(x) = {\sqrt{m} \over B} {\partial \varphi_{\pm}(x)
\over \partial x}.
\end{equation}
The propagator of $\varphi_\pm$ is obtained from the Eq.~(\ref{quadL})
\begin{equation}
\langle \varphi_\pm(x) \varphi_\pm(x') \rangle = - \ln (x - x').
\label{propagator}
\end{equation}
Eq.~(\ref{commrel}) gives the commutation relations for $\varphi_\pm$
\begin{equation}
[\varphi_{\pm}(x), \varphi_{\pm}(x')]=\pm i \pi \mbox{sign}(x-x').
\end{equation}
\par
The charge density at the upper and lower chiral boundaries are measured by
$( B / 2\pi m) y_{\pm}(x)$. Thus an operator $V(x)$ carrying the upper
and lower chiral charge $Q_{\pm}$ must satisfy
\begin{equation}
 [y_{\pm}(x), V(x')] = {2 \pi m \over B} Q_{\pm} \delta(x-x') V(x').
\end{equation}
Statistics of $V(x)$ is defined by the exchange phase $\theta$
\begin{equation}
V(x) V(x') = \exp(i \theta) V(x') V(x).
\end{equation}
If $V(x)$ is  fermionic, $\theta$ is quantized as $(2l+1)\pi$ for
integer-valued $l$.
Let us consider  operators of the form
\begin{eqnarray}
V_{p,q}(x)&=&:\exp(-ip \sqrt{m}(\varphi_+(x) - \varphi_-(x))/2
      \nonumber \\
      &-&iq (\varphi_+(x) + \varphi_-(x))/\sqrt{m}):.
 \label{vertex}
\end{eqnarray}
Normal ordering is taken with respect to the chiral boson vacuum.
It is easily seen that the chiral charge of this operator is
$Q_{\pm}=(p /2) \pm (q /m)$ and the statistical phase is $\theta=2\pi pq$.
We have assumed that the elementary excitations are fermions of unit charge.
Then $p$ must be an integer and statistics $2 \pi pq$ must be odd- or even-
intger multiples of $\pi$ for odd or even $p$ respectively.
Therefore we find that $p$, $q$ satisfies the selection rule $q=p/2 + {\bf Z}$.
This agrees with the selection rule derived in ~\cite{kawakamiyang}.
Physical meaning of the operators $V_{p,q}$ is as follows.
Chiral charge quantum numbers $Q_{\pm}$ imply that the operator $V_{p,q}(x)$
adds $p$ fermions and transfers $q$ units of quasi-hole carrying a fractional
charge $-1/m$ from one boundary to the other. As such state created by this
operator carries the momentum $2 q k_F$. (If one quasi-hole is transferred
>from one boundary to the other the momentum is changed by one $m$-th of that
for one electron transfer. One electron transfer from one boundary to the
other changes momentum $2m k_F$ since the maximum (minimum) momentum of the
fractional quantum Hall state is $ \pm m k_F$.) We also obtain the scaling
dimension
\begin{equation}
h_{p,q}={p^2 m \over 4}+{q^2 \over m}
\label{dimension}
\end{equation}
>from the Hamiltonian Eq.~(\ref{hamiltonian}) or, equivalently, from the
operator product expansion
\begin{equation}
\langle V(x) V^\dagger (x') \rangle = (x - x')^{-2 h_{p,q}}.
\end{equation}
The spectrum, the commutation relations and the Hamiltonian structure
indicate that the hydrodynamic collective field thoery is a  $c=1$ conformal
field theory. The dynamical degrees of freedom is a compact boson
$\varphi=\varphi_+ + \varphi_-$ of radius $\sqrt{m}$.
The spectrum also exhibits particle-hole duality interchanging
$\sqrt m \leftrightarrow 1 / \sqrt m$ at least in the thermodynamic
limit~\cite{reywu}.
\par
We now investigate fermion correlations along the boundaries of the FQH
droplet. The fermion correlation function (one-particle density matrix)
is defined by
\begin{equation}
\langle \psi^{\dagger}(x) \psi(x') \rangle =
\sum_{\mbox{int}} \langle 0|\psi^{\dagger}(x)|\mbox{int}\rangle
                 \langle \mbox{int}| \psi(x') |0 \rangle,
                 \label{int}
\end{equation}
where $|\mbox{int} \rangle$ denotes intermediate states of charge $-1$ and
fermionic statistics. It is easy to see that such states are created by
operators \cite{foot} as
\begin{equation}
|l \rangle = V_{-1,l+1/2} |0 \rangle, \ \ \ l \in \mbox{{\bf Z}}.
\label{lvertex}
\end{equation}
This state carries momentum $(2l+1)k_F$, statistics $\theta=(2l+1) \pi$ and
scaling dimension
\begin{equation}
h_{-1,l+1/2}={1 \over 4}\left( m+{(2l+1)^2 \over m} \right).
\end{equation}
Since
\begin{eqnarray}
V_{-1,l+1/2} &=&
    :\exp[-{i \over 2}(\sqrt{m}+{2l+1 \over \sqrt{m}})\varphi_{+}(x)
          \nonumber \\
          &+& {i \over 2}(\sqrt{m}-{2l+1 \over \sqrt{m}})\varphi_{-}(x)]:
\end{eqnarray}
the state $|l \rangle $ consists of a mixture of the two boundary states of
different chiralities $\varphi_{+}(x), \varphi_{-}(x)$.
\par
We are interested in the behavior of the correlation function as the
width of the FQH droplet is continuously changed.
When the droplet width is wider than the magnetic length $\sim 1 / \sqrt B$,
the two boundary states of opposite chiralities are decoupled each other.
Therefore the intermediate states in (\ref{int}) must be created only by
one of the chiral bosons $\varphi_{+}(x)$ or $ \varphi_{-}(x)$ at each
edges. Unique choices of the appropriate operators in (\ref{lvertex}) are
\begin{equation}
V_{-1,m/2}=:\exp[-i \sqrt{m} \varphi_{+}(x)]
\end{equation}
and
\begin{equation}
V_{-1,-m/2}=:\exp[+i \sqrt{m} \varphi_{-}(x)].
\end{equation}
Therefore for FQH droplet with widely separated boundaries the correlation
function is given by
\begin{equation}
\langle \psi^{\dagger}(r) \psi(0) \rangle \sim r^{-m} \cos (m k_F r)
+ \cdots.
\label{corrLaugh}
\end{equation}
This agrees with the edge state correlation function of the FQHE obtained
previously by Wen~\cite{wen}.
As shown in the second paper in \cite{iso} the 2-d fermion correlation function
in the LLL and in the Landau gauge is expressed in terms of the 1-d fermion
correlation as
\begin{eqnarray}
\langle \psi_0^{\dagger} (z, \bar{z}) \psi_0(z',\bar{z'}) \rangle
&\sim&
 e^{-(B/2\hbar)[(y-i\hbar\partial_r/B)^2
+(y'-i\hbar\partial_r/B)^2]}
 \nonumber \\
 && \times \langle \psi^{\dagger} (r) \psi(0) \rangle
\end{eqnarray}
where $\psi_0(z, \bar{z})$ is the LLL fermion operator.
It is clear that the 2-d correlation in the long distance
limit shows a power-law behavior only when $y$ and $y'$ lie on either one
of the two boundaries $y=y'=\pm m k_F / B$ of the FQH droplet.
As one moves into the bulk from the boundaries, the 2-d correlation is
exponentially damped.
\par
Next consider the limit in which two edges of the FQH droplet are close
each other of order of the magnetic length. We assume that the edge state
fluctuations remain small so that topology change of the droplet does not
take place. In this limit the two edge states with different chiralities
interact strongly each other. Accordingly all the states with the correct
quantum numbers $|l \rangle$ can contribute to the intermediate states in
Eq.~(\ref{int}). The long-distance correlation is given by the most relevant
operator. Such an operator with the smallest scaling dimension is
$V_{-1,\pm 1/2}$. Since this operator has momentum  $k_F$ and scaling
dimension $h_{-1,\pm 1/2}=(m+1/m)/4$, the fermion correlation function becomes
\begin{equation}
\langle \psi^{\dagger}(r) \psi(0) \rangle \sim r^{-(m+1/m)/2} \cos ( k_F r).
\label{corrCS}
\end{equation}
This is the Tomonaga-Luttinger liquid correlation behavior with
$\cos ( k_F r)$ oscillation. The exponent is the same as  the one found in
\cite{kawakamiyang} for the CS model. In deriving Eq.~(\ref{corrCS}) we have
assumed that there are no further mixing between the left- and the
right-handed edge states by, for instance, impurities on the edges.
Therefore we have shown that the fermion correlation function along the
boundaries of the FQH droplet reproduces that of the chiral Luttinger
liquid of the edge states and interpolates to the correlation function of
the CS model as the droplet width is continuously narrowed.
\par
Density correlation function is derived similarly.
When the droplet width is wide the density correlations at either boundaries
are
\begin{equation}
\langle y_{\pm}(x) y_{\pm}(x') \rangle =
{m \over B^2} \partial_x \partial_{x'}
\langle \varphi_{\pm}(x)  \varphi_{\pm}(x') \rangle
= {m \over B^2} {1 \over (x-x')^2}.
\end{equation}
Once the droplet width becomes narrow, the otherwise decoupled chiral
edge excitations begin to mix each other.
Since the state $\rho(x)|0 \rangle$ has charge $0$ and bosonic
statistics, intermediate states in $\langle \rho(x) \rho(x') \rangle$
are created by operators $V_{0,l}$ for integer-valued $l$.
This operator has
$ Q_{\pm}=\pm(l/ m), \theta=0,  h_{0,l}=l^2 /m$
and momentum $2l k_F$.
Therefore the density correlation function of the narrow FQH droplet is
given by
\begin{equation}
\langle \rho(r) \rho(0) \rangle \sim A_0 r^{-2} +\sum_{l = 1}^\infty
A_{l} \  r^{-2l^2 /m} \cos(2l k_F r)
 \end{equation}
and exhibits the expected $\cos(2 l k_F r)$ oscillations.
\par
In this Letter, applying the hydrodynamic collective field theory approach to
the FQH droplet dynamics, we have shown that the correlation
functions of the FQHE edge state and the Calogero-Sutherland model are related
each other.
The 1-d representation of the Laughlin wave fucntion Eq.~(\ref{Laughlin})
describes both the edge state correlation and the correlation of the
Calogero-Sutherland model.
When the width of the droplet is larger than the magnetic length
the fermion correlation behaves as
$\langle \psi^{\dagger}(r) \psi(0) \rangle \sim r^{-m} \cos(m k_F r) $
while in the narrow limit it behaves as
$\langle \psi^{\dagger}(r) \psi(0) \rangle \sim r^{-(m+1/m)/2} \cos(k_F r) $.
Numerical calculation for the adiabatic interpolation of the fermion
correlation and detailed exposition of this Letter will be reported
separately \cite{numerical}.
\par
 S.I. acknowledges warm hospitality of the Center
 for Theoretical Physics at Seoul National University, where this work was
 initiated. S.J.R. thanks A. Polychronakos for useful discussions. S.J.R. was
supported in part by KOSEF-SRC program, Ministry of
 Education BSRI-94-2418 and KRF-Nondirected Research Grant`93 (SJR).


\begin{thebibliography}{1}

\bibitem{calogero} F. Calogero, J. Math. Phys. {\bf 10} (1969) 2197;
  J. Math. Phys. {\bf 12} (1971) 418.

\bibitem{sutherland}
B. Sutherland, J. Math. Phys. {\bf 12} (1971) 246, 251; Phys. Rev. {\bf A4}
(1971) 2019; {\bf A5} (1972) 1372.

\bibitem{anyon}
A.P. Polychronakos, Nucl. Phys. {\bf B324} (1989) 597;
D. Bernard and Y. S. Wu, cond-mat/9404025 preprint, 1994 (unpublished).

\bibitem{haldane}
H.D.M. Haldane, cond-mat/9401001 Princeton preprint, 1994 (unpublished).

\bibitem{simon}
B. D. Simon, P. A. Lee and B. A. Altshuler, Phys. Rev. Lett. {\bf 70} (1993)
 4122; Phys. Rev. Lett. {\bf 72} (1994) 64.

\bibitem{poly}
J.A. Minahan and A.P. Polychronakos, CERN-TH.7243/94, hep-th/9404192
preprint, 1994 (unpublished); F. Lesage, V. Pasquier and D. Servan,
hep-th/9405008 Saclay preprint 1994 (unpublished); Z.N.C. Ha,
IASSNS-HEP-94/27 preprint, 1994 (unpublished).

\bibitem{kawakami}
N. Kawakami, Phys. Rev. Lett. {\bf 71} (1993) 275; J.Phys.Soc.Jpn {\bf 62}
(1993) 2270, 2419; cond-mat/9402011, 1994 (unpublished).

\bibitem{W}
S. Iso, D. Karabali and B. Sakita, Phys. Lett. {\bf B296} (1992) 143;
A. Cappeli, C.A. Trugenberger and G.R. Zemba, Nucl. Phys. {\bf B396} (1993)
465;
D. Karabali, SU-4240-580 preprint, 1994 (unpublished).

\bibitem{wadati}
K. Hikami and M. Wadati, UT/W93-012 preprint , 1993 (unpublished).

\bibitem{iso}
H. Azuma and S.Iso, UT-660, hep-th/9312001 preprint, 1993 (unpublished);
  S. Iso, UT-676, cond-mat/9404075 preprint, 1994 (unpublished).

\bibitem{wen}
X.-G. Wen, Phys. Rev. Lett. {\bf 64} (1990) 2206;
Phys. Rev. {\bf B41} (1990) 12838; {\bf B43} (1991) 11025; {\bf B44}
(1991) 5708.

\bibitem{kawakamiyang}
N. Kawakami and S-K. Yang,  Phys. Rev. Lett. {\bf 67} (1991) 2493.
\bibitem{stonefisher}
M. Stone and M. P. A. Fisher, NSF-ITP-94-15 preprint, 1994 (unpublished).

\bibitem{iks}
S.Iso, D.Karabali and B.Sakita, Nucl.Phys. {\bf B388} (1992) 700.

\bibitem{girvin}
S. M. Girvin and H. MacDonald, Phys.Rev.Lett. {\bf 58} (1987) 1252;
S. C. Zhang, T. H. Hanson and S. Kivelson, Phys.Rev.Lett. {\bf 62} (1989) 82;
N. Read, Phys.Rev.Lett. {\bf 62} (1989) 86.


\bibitem{gjgmp}
S.M. Girvin and T. Jach, Phys. Rev. \bf B29 \rm (1983) 5617; S.M. Girvin,
A.H. MacDonald and P.M. Platzman, Phys. Rev. \bf B33 \rm (1986) 248;
B. Sakita, D.-N. Sheng and Z. -B. Su, Phys. Rev. {\bf B44} (1991) 11510.

\bibitem{reywu}
S.-J. Rey and Y.-S. Wu, to appear.

\bibitem{foot}
Of course neutral boson operators may be added to the intermediate staets.
As they are less relevant in the  long-distance limit we have neglected them
in our calculation.

\bibitem{numerical} S. Iso and S-J. Rey, to appear; \\
Yoshioka and Ogata have also discussed the fermion correlations in the
numerical calculation and is consistent with our result;
D. Yoshioka and M. Ogata, J. Phys. Soc. Jpn. {\bf 62} (1993) 2988.

\end{thebibliography}
\end{document}